\documentclass[aps,prd,groupedaddress,twocolumn]{revtex4}

\usepackage{amsmath,amssymb,bm}

\usepackage{graphicx}
\usepackage{amsfonts}
\usepackage{amssymb}
\usepackage{amsbsy}
\usepackage{amsmath}
\usepackage{latexsym}
\usepackage{bm}
\usepackage{color}
\usepackage{subfigure}
\usepackage{comment}

\def\p {\partial}

\def\be {\begin{eqnarray}}
\def\ee {\end{eqnarray}}
\def\nn {\nonumber}

\begin{document}

\title{Quantum gravitational collapse as a Dirac particle on the half-line}

\author{Syed Moeez Hassan}
\email[]{shassan@unb.ca}
\author{Viqar Husain}
\email[]{vhusain@unb.ca}
\author{Jonathan Ziprick}
\email[]{jziprick@unb.ca}

\affiliation{University of New Brunswick\\
Department of Mathematics and Statistics\\
Fredericton, NB E3B 5A3, Canada}


\begin{abstract}
 
We show that the quantum dynamics of a thin spherical shell in general relativity is equivalent to the Coulomb-Dirac equation on the half line. The Hamiltonian has a one-parameter family of self-adjoint extensions with a discrete  energy spectrum $|E| < m$, and a continuum of scattering states for $|E|>m$, where $m$ is the rest mass of the shell and $E$ is the Arnowitt-Deser-Misner mass. For sufficiently large $m$, the ground state energy level is negative. This suggests that classical positivity of energy does not survive quantization. The scattering states provide a realization of singularity avoidance. We speculate on the consequences of these results for black hole radiation. 
 
\end{abstract}

\maketitle

\section{Introduction}

The study of singularity formation in general relativity, and its potential avoidance in a quantum theory of gravity, has been the focus of  numerous papers over the past few decades. This is a topic of singular importance for foundational physics as it applies to black holes and cosmology in  extreme gravitational fields.  

With no accepted quantum theory of gravity, all studies of singularity avoidance use simpler systems derived from general relativity, or other gravity theories. Such systems arise by imposing symmetries on the spacetime metric, and typically fall into two classes: the homogeneous cosmological models (also known as mini-superspace reductions), where the gravitational field equations are classically reduced to particle mechanics, or to spherically symmetric asymptotically flat models.   

There are numerous papers on the quantization of homogeneous cosmologies beginning in the 1970s \cite{Misner:1969hg, Ryan:1975jw}, to their present incarnation in loop quantum cosmology (see for example \cite{Agullo:2016tjh} for a recent review). But to date, there is no complete quantization of spherically symmetric gravitational collapse in general relativity \cite{Husain:2009vx, Gambini:2012zz}; the difficulty is that  these models with matter fields are 2-dimensional (non-conformal) field theories.

With few exceptions, all models have been quantized as scalar particles; the Wheeler-DeWitt operator does not admit a general Dirac square root quantization due to the fact that gravitational momenta and configuration variables are mixed in its kinetic part, unlike  that of standard particle Hamiltonians. The case of homogeneous isotropic cosmology is an exception, where after a time gauge fixing the resulting physical Hamiltonian  does permit the Dirac square root  \cite{Kreienbuehl:2009ub}. 

Here we describe a complete Dirac square root quantization of an asymptotically flat gravitational collapse model,  the thin spherical shell in general relativity. This is amongst the simplest  models, and may be viewed as the ``hydrogen atom" for gravitational collapse. This model, and others like it, are well-studied classically \cite{Poisson:2004,Adler:2005vn}.  A  comprehensive analysis of Wheeler-DeWitt quantization appeared in \cite{Hajicek:1992}; this last paper maps the problem to the ``scalar Hydrogen" atom with the Klein-Gordon inner product. Recently the same problem was studied using the polymer quantization method \cite{Ziprick:2016ogy}. The common feature of both these papers is that the quantization method is that of scalar particles. Our treatment maps the model to the radial 2-spinor hydrogen atom.

In the following  we first recall the shell collapse model and present the Hamiltonian formulation we use for quantization. We then   describe in detail the construction of the spinor Hamiltonian as a self-adjoint operator on the 2-spinor Hilbert space, and compute  its spectrum. We conclude with a discussion of our main results.  (We work throughout in units where $G=c=\hbar=1$.)

 \section{Gravitational dynamics of the thin shell}
 
The dynamics of the thin spherical shell is a simple model for gravitational collapse in general relativity. It is constructed by patching an exterior Schwarzschild metric to an interior Minkowski metric along a timelike surface $\Sigma$ with topology $ \mathbb{R}\times S^2$.  The interface between these metrics is called the thin shell. Its dynamics is obtained from the junction conditions. These  require that at the interface (i) the metric is continuous, and (ii) the jump in the extrinsic curvature equal the shell's stress-energy tensor \cite{Poisson:2004,Adler:2005vn}.  

 The interior, surface and  exterior metrics are respectively 
 \be
 ds_{int}^2 &=& -dt^2 + dr^2 + r^2 d\Omega^2\ , \nn\\
 d\sigma_\Sigma^2 &=& -d\tau^2 + X^2(\tau)d\Omega^2\ ,\nn\\
 ds_{ext}^2 &=&  - f(R)dT^2 + f^{-1}(R) dR^2 + R^2 d\Omega^2. 
 \ee
 The spherical coordinates $(\theta, \phi)$ are the same in the three regions, $\tau$ is the proper time of the shell, $X(\tau)$ is the shell's radius (the configuration variable of the model), $f(R)=1-2E/R$ and the exterior Schwarzschild parameter $E$ is the total gravitational mass of the shell as seen at infinity.  The junction condition on the metrics at $\Sigma$ are 
 \be
 g_{int}|_\Sigma = \sigma_\Sigma = g_{ext}|_\Sigma .
 \ee
  These give  $X(\tau) = r(\tau)=R(\tau)$, and 
 \be
  \dot{t}^2 -\dot{r}^2 = 1 = f \dot{T}^2 - f^{-1}\dot{R}^2. \label{junc1}
 \ee
 where a dot denotes the derivative with respect to $\tau$.  
 
 The shell stress energy tensor is taken to be that of pressureless dust $T_{ab} = \sigma u_au_b$, which is confined to the boundary between the interior Minkowski metric and the exterior Schwarzschild metric.  To derive $T_{ab}$, the jump in the extrinsic curvature from the interior Minkowski metric to the exterior Schwarzschild metric, is computed using the outward pointing normals 
 $n^{int}_a dx^a = -\dot{r}dt + \dot{t}dr$ and $n^{ext}_adx^a = -\dot{R}dT + \dot{T}dR$. 
 This leads to the result  \cite{Poisson:2004}   
 \be
  4\pi X^2(\tau) \sigma  = \text{constant} \equiv m
 \ee
 and 
 \be
E = m\sqrt{1+ \dot{X}^2}-V, \quad\quad V \equiv \frac{m^2}{2X}. \label{H1}
\ee
This is the final dynamical equation for the thin shell. It also provides  an expression for the Hamiltonian $H(X,\dot{X};m)\equiv E$ as a function of the shell's mass, $m$, its radius $X(\tau)$, and velocity $\dot{X}$ with respect to shell proper time $\tau$. The Hamiltonian is just the Arnowitt-Deser-Misner (ADM) mass of the exterior  Schwarzschild solution.

For  quantization there is a more convenient form of $H$  that uses the interior Minkowski time $t$ rather than the proper time $\tau$. This is obtained from (\ref{H1}) using  the first equation in  (\ref{junc1}), which gives 
\be
\label{eom}
H\equiv E = \frac{m}{ \sqrt{1-(dX/dt)^2}} - V  . \label{H2}
\ee
Taking the shell radius $X(t)$ as a phase space variable, its conjugate momentum $P$ is obtained using the Hamilton equation  
\be
P = \int dH \left(\frac{dX}{dt}\right)^{-1} = \int dH \left(1- \displaystyle \frac{m^2}{(H+V)^2}\right)^{-\frac{1}{2}}.
\ee
This gives a quadratic expression for the Hamiltonian which we shall use for quantization:
\be
\label{constraint}
\left(H + V\right)^2 = P^2+m^2 . \label{H}
\ee
 
\section{Quantization using the Dirac square root}

A scalar-particle quantization  starting from  (\ref{constraint}) was studied for the electromagnetic case (ie. with potential $V=Ze^2/R$) in  \cite{Herbst:1977a} where the square root of the Hamiltonian (\ref{H}) was used, and  in \cite{Hajicek:1992, Ziprick:2016ogy} where it was treated as a Wheeler-DeWitt equation. Here we describe a new quantization which involves taking the square root ˆ\`a la Dirac, and defining a  matrix Hamiltonian operator in a 2-spinor Hilbert space. 

 At the algebraic level we have  $[X,P]=i$, and we propose the following Hamiltonian, using the $2\times 2$ identity $I$ and the Pauli matrices
 \be
 H I = P \sigma_3 +m\sigma_2 - V(X) I =
\begin{bmatrix}
  P - V       &  -im     \\
     im    &  -P-V  
\end{bmatrix}.
 \ee   
 It is immediate that this yields $(H + V)^2 = P^2 +m^2$.  With this prescription the quantization of the thin shell dynamics turns out to be the Dirac equation on the half line for a Coulomb potential with charge given by the rest mass $m$.  

\subsection{Hilbert space and Hamiltonian operator}

To realize $H$ as a self-adjoint operator on a Hilbert space, let us consider the space of 2-spinor functions  
  \be
\Phi \equiv \begin{bmatrix}
    \phi_1     \\
    \phi_2 
\end{bmatrix}
 \ee
with inner product 
\be
\label{ip}
( \Psi , \Phi ) := \int_0^\infty dX \ \Psi^\dagger \Phi = \int_0^\infty dX \left(\psi_1^* \phi_1 + \psi_2^* \phi_2\right),
\label{IP}
\ee
where $\Psi^\dagger\equiv \Psi^{T*}$, and the standard representation
\be
\hat{X} \phi = X\phi, \qquad \qquad \hat{P} \phi = -i \partial_X \phi.
\ee
The expression for the Hamiltonian operator on this Hilbert space is  
\be
\label{Hop}
 \hat{H}  =
\begin{bmatrix}
  -i\p_X - V       &  -im     \\
     im    &  i\p_X -V  
\end{bmatrix}.
 \ee
A definition of $\hat{H}$ as a self-adjoint operator requires a precise specification of its domain. Two conditions must be met for this: that its adjoint $\hat{H}^A$ have the same algebraic expression, i.e. that $\hat{H}$ is symmetric, and secondly that $\hat{H}$ and its adjoint have identical  domain, i.e. ${\cal D}(\hat{H}) =  {\cal D}(\hat{H}^A)$. Let $\Phi \in {\cal D}(\hat{H})$ and $\Psi \in  {\cal D}(\hat{H}^A)$.  Then
 \be
\label{sym}
(\Psi, \hat{H} \Phi)  
  = (\hat{H} \Psi, \Phi) + i \left[ \psi_1^* \phi_1 - \psi_2^* \phi_2 \right]^\infty_0.
\ee
Since the wave functions are required to fall off sufficiently fast at infinity for square integrability with the inner product $(\ref{IP})$, the condition that $\hat{H}$ is symmetric requires  
\be
\lim_{X\rightarrow 0}  \left[ \psi_1^* \phi_1 - \psi_2^* \phi_2 \right] =0, \label{surface}
\ee
and that  $\hat{H}$  be self adjoint requires identical boundary conditions for  
${\cal D}(\hat{H}) $ and  $ {\cal D}(\hat{H}^A)$.  To address these  conditions let us turn to the eigenvalue problem for the proposed  Hamiltonian (\ref{Hop}). 

Writing $\p_X\phi \equiv \phi'$, the eigenvalue equations are 
 
 \begin{subequations}
\label{e-value}
\be
  -i \phi_1'-V \phi_1 - im \phi_2  &=& E \phi_1, \\
  i \phi_2' -V \phi_2 + im \phi_1 &=& E \phi_2  .
  \ee
The leading order forms of $\phi_1, \phi_2$ as $X \to 0$  are
\end{subequations}
\be
\label{smallx}
\phi_1 \sim X^{im^2/2},  \quad \phi_2 \sim  X^{-im^2/2}, \quad \  X  \ll  1.
\ee
These are compatible with (\ref{surface}), a point which we elaborate below.   

For the full problem it is convenient to define  $\phi_1 \equiv X^{i\frac{m^2}{2}} f(X)$ and $\phi_2 \equiv  X^{-i \frac{m^2}{2}} g(X)$. 
With this,  the pair of equations (\ref{e-value}) may be written as 
\be
\label{feq}
f + \frac{X^{-im^2}}{m} \left(iEg + g' \right) &=& 0, \\
\label{geq}
g'' - \frac{im^2}{X}g' + \left(k^2 + \frac{m^2 E}{X} \right)g &=& 0, \label{g-eqn}
\ee
where $k^2=E^2-m^2$.   The second equation (\ref{geq}) for  $g$ may be transformed  to Kummer's equation \cite{NIST:DLMF},
\be
z \frac{d^2 w}{dz^2} + (b-z) \frac{dw}{dz} - a w = 0
\ee
by writing $g(X) = e^{-ikX} X^{im^2 + 1} w(X)$, with $z = 2ikX$, and defining the constants 
\be
a &\equiv& 1+ \frac{im^2}{2} \left( 1 + \frac{E}{k} \right), \quad b \equiv 2+ im^2.\nn 
\ee
The solution of the eigenvalue equations may then be written as      
 \be 
 \label{phii}
\phi_2 &=& \left[c_1 M(a, b;z) + c_2 U(a, b; z) \right] e^{ikX} X^{1+\frac{im^2}{2}}, \\
 \phi_1 &=& - \frac{1}{m}\left[ \phi_2' + i \phi_2\left( E +V \right)\right],   
\ee
where $M(a, b, z)$ and $U(a, b, z)$ are the two independent solutions to Kummer's equation \cite{NIST:DLMF} and $c_1, c_2 \in \mathbb{C}$  are arbitrary constants.
 
For $X \ll 1$ the eigenfunctions take the form
\be
\phi_2 &=&  c_2  (2ik)^{1-b} \frac{\Gamma(b-1)}{\Gamma(a)}) X^{- im^2/2} + {\cal O}(X), \\
\phi_1 &=&  \frac{1-b}{m}  \left( c_1 + c_2 \frac{\Gamma(1-b)}{\Gamma(a-b+1)} \right) X^{im^2/2} \nn \\
&& + {\cal O}(X).
\ee

Thus for $X \ll 1$  
\be 
\label{ratio}
 \frac{\phi_1}{\phi_2} &=& X^{im^2} \rho_m(E), \\
 \rho_m(E) &\equiv& \frac{(1-b)\Gamma(a) (2ik)^{b-1}}{m\Gamma(b-1)} \nn \\
 && \times  \left[ \eta + \frac{\Gamma(1-b)}{\Gamma(a+1-b)}\right] 
\ee
where $\eta \equiv c_1/c_2$.  With the condition (\ref{ratio})  also imposed on $\Psi \in {\cal D}(\hat{H}^A)$, the surface term (\ref{surface}) vanishes provided 
\be
\label{ev-condition}
\rho_m(E) = e^{i\theta} 
\ee 
where $\theta \in [-\pi,\pi)$  specifies a one-parameter family of self-adjoint extensions of $\hat{H}$.  {\it This is also the eigenvalue equation} for a given $m$, and a choice of $\theta$. Its solution $E=E(m,\theta)$ in general requires $\eta=\eta(E,m)$.  

The above steps  complete the specification of the Hilbert space ${\cal H}$:
\be
\mathcal{H} = && \left\{ \begin{bmatrix}
    \phi_1     \\
    \phi_2 
\end{bmatrix} \in \mathcal{L}^2(\mathbb{R}^+) \oplus \mathcal{L}^2(\mathbb{R}^+) \right. \nn \\
 &&\left.  |  \lim_{X\rightarrow 0} \left( \frac{\phi_1}{\phi_2} - e^{i(\theta + m^2 \ln X)} \right) =0  \right\}  
 \label{H-space}
\ee
The origin is excluded from the domain of the functions because the   solutions (\ref{smallx}) oscillate rapidly as $X\rightarrow 0$ so $\phi_1(0)$ and $\phi_2(0)$ are not defined.  It is readily verified that the standard probability and energy-momentum currents 
\be
j^\mu(\Phi) = \Phi^\dagger \gamma^0 \gamma^\mu \Phi, \qquad  p^\mu(\Phi) = \Phi^\dagger \gamma^0 \gamma^\mu \hat{H} \Phi,  
\ee
follow from the time dependent Dirac equation for all $\Phi\in {\cal H}$, with $\gamma^0:= \sigma^2$ and $\gamma^1:=-\sigma^1$. (We mention this elementary  point because in the scalar-particle quantization of the model  \cite{Hajicek:1992, Ziprick:2016ogy}, the probability density is not positive definite due to the Klein-Gordon inner product.) 

In concluding this section let us note that the specification of the Hilbert space as in (\ref{H-space}) arises due to the starting choice of inner product (\ref{IP}). The same choice was used in  \cite{Hajicek:1992} where this model was quantized using the Wheeler-DeWitt equation; one of these authors has shown \cite{Hajicek:1992mx} that the measure $4\pi X^2 dX$ yields a unitarily equivalent theory. A similar analysis applies here as well, and so we work with the more simple choice of measure.

\section{Spectrum of the Hamiltonian}

To determine the spectrum, let us first consider the eigenfunctions (\ref{phii}) which hold for $E\ne m$. (The $E=m$ case is considered separately below.) Their behaviour for  $X\gg 1$ is given by the linear combination  \cite{NIST:DLMF}
\be
\label{phi_large}
\phi_2(X) &\approx& \frac{1}{(2ik)^{a}} \bigg[c_1 \frac{(-1)^a \Gamma(b)}{\Gamma(b-a)} +c_2  \bigg] e^{ -i\left(kX+\frac{Em^2}{2k}\ln X\right)} \nn \\
&+& \bigg[c_1 \frac{\Gamma(b)}{\Gamma(a)} (2ik)^{a-b} \bigg] e^{i\left(kX+\frac{Em^2}{2k}\ln X\right)}.
\ee
This expansion is useful for the following cases.

\begin{figure}
\subfigure[\label{m=0.5}$\ m=0.5$]
{\includegraphics[scale=0.40]{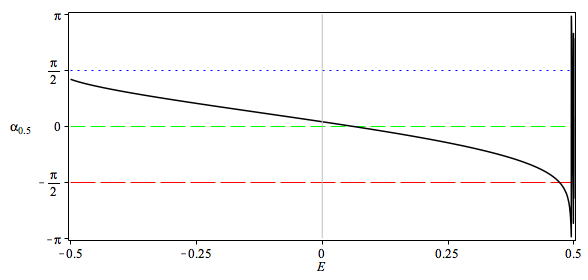}}\\
\subfigure[\label{m=1}$\ m=1.0$]
{\includegraphics[scale=0.40]{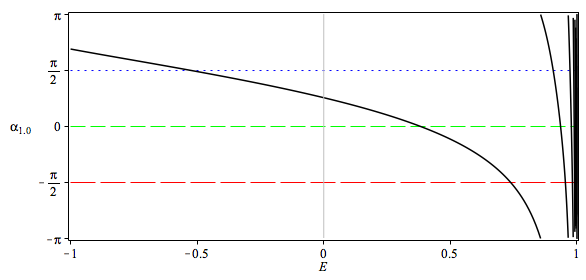}}\\
\subfigure[\label{m=2.0}$\ m=2.0$]
{\includegraphics[scale=0.40]{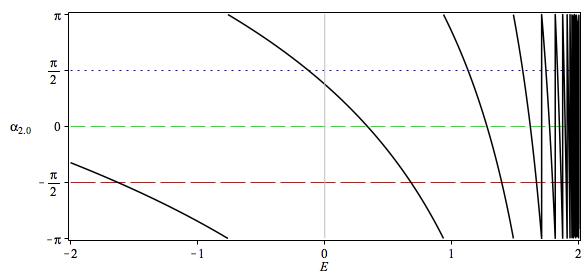}}\\
\subfigure[\label{m=2.6}$\ m=2.6$]
{\includegraphics[scale=0.40]{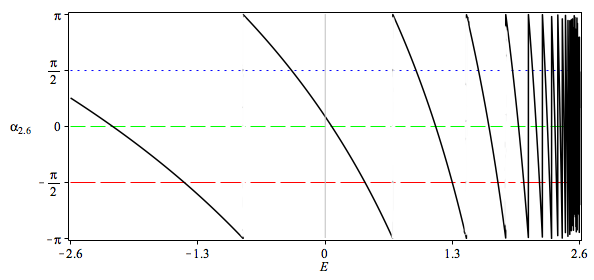}}
\caption{Plots of the bound state spectrum, $\alpha_m$ vs. $E$, for the permitted range $E\in (-m,m)$. The horizontal lines are $\theta =0,\pm\pi/2$; their intersections with $\alpha_m$ give the spectral values. The clustering of eigenvalues near $+m$ is evident in all cases.  As $m$ increases the lowest eigenvalue moves toward $-m$ for all $\theta\in [-\pi,\pi)$. The last graph shows that  the ground state energy is negative for any $\theta$. These levels correspond to the type of classical orbits shown in Fig. 2}
\end{figure}

  \underline{$|E|<m$}: In this case $k$ is purely imaginary and normalizability requires exponential damping at large $X$. Therefore we must have  $k=-i\sqrt{m^2-E^2}$.  With this, it follows that the l.h.s. in (\ref{ev-condition}) has unit magnitude provided $\eta = 0$ (using the identity $\overline{\Gamma(t)} = \Gamma(\bar{t})$). The argument of $\rho_m(E)$ then gives the bound state spectrum according to (\ref{ev-condition}). This is illustrated in  Figs. \ref{m=0.5} to \ref{m=2.6}: the eigenvalues  are those values of $E$ for which $\alpha_m(E)\equiv\arg(\rho_m(E)) = \theta$.  

Fig. \ref{m=0.5} shows the spectrum  for $m=0.5$. It is evident that  eigenvalues accumulate very close to $E=m$ for all $\theta\in [-\pi,\pi)$. Negative ground state energies exist  for finite ranges of $\theta$,  while all other eigenvalues are positive. Fig. \ref{m=1} shows an instance of a negative value of ground state energy for with $\theta \in (0.8, 2.1)$ (approximately), while other $\theta$ choices give positive values. 
  
The graphs demonstrate that in all cases, there are an infinite number of eigenvalues for $E\in(-m,m)$, with the gap between eigenvalues decreasing as the energy increases.  Furthermore, as $m$ decreases the eigenvalues all move closer to the upper limit $+m$. As $m$ increases, the lower end of the spectrum moves toward $-m$, and the number of negative eigenvalues increases.  Importantly, as demonstrated in Fig. \ref{m=2.6}, {\it for sufficiently large $m$, the ground state energy is negative for any choice of $\theta$.}   
 
 \underline{$|E|>m$}: In this case $k\in \mathbb{R}$, and we can see from the large X behaviour (\ref{phi_large}) that
$\phi_2$ and $\phi_1$ are linear combinations of ingoing and outgoing modes.  Furthermore, it is always possible to find algebraically an $\eta(E,m)\ne0$  such that the l.h.s.  of (\ref{ev-condition}) has magnitude unity. These are the scattering states.

The classical correspondence of these states  is evident from (\ref{H2}): for $X\rightarrow \infty$ classical trajectories for $E>m$ have finite outgoing or ingoing velocities $\displaystyle \dot{X} = \pm \sqrt{1-m^2/E^2}$.  The former describe shells with sufficient energy to escape to infinity, whereas the ingoing ones come in from infinity with finite velocity and form a black hole. These are unlike the trajectories corresponding to the bound states, where the shell trajectories never reach infinity. 
  
  \underline{$E=\pm m$}: In this case the $X\ll 1$ limiting forms are still given by (\ref{smallx}), however  (\ref{g-eqn}) reduces to the Bessel equation, leading to the solution 
  \be
  \phi_2 &=& \sqrt{X} \left(c_1 J_\nu(2\sqrt{m^3X})  + c_2 Y_\nu(2\sqrt{m^3X})\right)    
  \ee
where  $\nu = -(1+im^2)$, and $J_\nu, Y_\nu$ are the Bessel functions. The large $X$ behaviour of these functions are oscillatory with $X^{-1/4}$ damping. However, with the $\sqrt{X}$ prefactor, the oscillations grow in amplitude as $X^{1/4}$.  This means that the solution for this case must vanish since it diverges at large $X$. Hence there is no $E=\pm m$ eigenvalue. 

\begin{figure}
\centering
\includegraphics[scale=0.40]{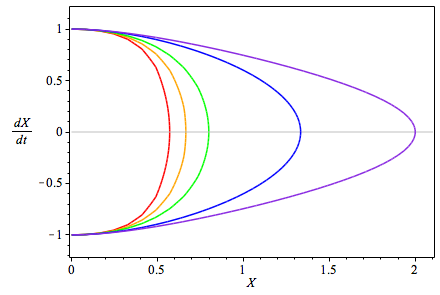}\\
\caption{Classical trajectories from Eqn. (\ref{H2}) for $m=2$ and $E=-1.5,-1.0,-0.5,0.5,1.0$, displayed innermost to outward. The first three correspond to the negative mass, and the last two to positive mass  exterior Schwarzschild solutions. The bound state spectrum corresponds to a selection from such classical orbits.}
\label{NegE}
\end{figure}

\section{Discussion}  

To summarize, we have shown that one of the simplest models for gravitational collapse in general relativity  permits a Dirac square root quantization.  The  Hamiltonian has a one parameter  family of self-adjoint extensions, which give a  discrete bound state spectrum $E\in (-m,m)$, and a continuum of scattering states for $|E|>m$. As we have demonstrated through Fig. 2, these states correspond to specific classical trajectories, similar for example, to the case of the hydrogen spectrum. 

Of particular note is that there is no value of $\theta$ that yields a spectrum {\it bounded below at zero for all }$m>0$. Thus, {\it this quantization generically gives quantum states with negative ADM mass.}  Furthermore, in comparison to the scalar-particle quantization in \cite{Hajicek:1992}, our approach has two distinct advantages: the probability density is positive definite, and the quantization is valid for all $m$. This is unlike the limitation $m<1$ in the scalar quantization in \cite{Hajicek:1992}. (On this last point, we note that the spectrum of the  Hamiltonian denoted $H_{13}= \sqrt{-\p^2/\p r^2+m^2} - m^2/2r$ in  \cite{Hajicek:1992} was derived for the electro-magnetic case in \cite{Herbst:1977a}; the theorems in this latter paper are for arbitrary mass $m$ but for potential $V=-Ze^2/r$ restricted to  $Ze^2<2/\pi$, where $Z \in \mathbb{N}$ and $e$ is the electron charge. Therefore, applied directly to the shell case, where $Ze^2=m^2/2$, the results of  \cite{Herbst:1977a} hold for $m^2<4/\pi$.)   Lastly, we note that the selection of a possible value of $\theta$ requires experimental input, as it does even for the Hydrogen atom ground state \cite{Fewster:1993yz}.  

The negative energy bound states  (for example levels such as  $E\approx -1.5$ for $m=2$)  may be associated to classical solutions arising from (\ref{H2}), shown in Fig. \ref{NegE}. The negative energy comes from the fact that the (negative) gravitational binding energy dominates the rest and kinetic energies of the shell. The exterior solution in these cases is negative mass Schwarzschild, nevertheless the spectrum is bounded below. For the positive energy eigenstates, it is possible to compare $\langle \hat{X} \rangle_\theta$ with the corresponding classical Schwarzschild radius $2E_\theta$. However, because  $\langle \hat{X}^2 \rangle_\theta>0$, it is unclear to us what physics such a comparison would yield, other than that there is a  bound state confined near $X=0$, which we already know. 

How does the spectrum we have found address ``singularity avoidance" in quantum gravity? This question may be answered by using the bound states on the one hand, and the scattering states on the other. In a cosmological context where the classical solution has a big bang, singularity resolution  is manifested as a ``bounce." This is because the setting is usually one in which there is no physical Hamiltonian for a matter-gravity system whose spectrum might be computed (however see \cite{Husain:2011tk,Husain:2011tm} for an exception). A quantum version of the Hamiltonian constraint is used to generate ``evolution'' with respect to a matter time (as for example the scalar field in \cite{Ashtekar:2006rx}), and one can then ask if such an evolution terminates.  

The scattering states for $E>m$ may thus be viewed as exemplifying the form of singularity avoidance usually discussed in the quantum gravity  literature: the physical picture is one of an incoming shell that bounces off the origin with a phase shift given by the spectral condition (\ref{ev-condition}) for  a given $E$.   There are of course no classical scattering solutions with initial ingoing velocity such that $E>m$: the classical shell comes in, crosses the Schwarzschild radius, and forms a black hole.  

Turning to the bound states, the manifestation of singularity avoidance in atomic systems is that the spectrum is bounded below.  This is also what we observe here. However, for the cases of  sufficiently large $m$, where the ground state energy is negative for any $\theta$, the external metric is negative mass Schwarzschild by classical correspondence. This leaves the unusual combination of ``singularity avoidance" as in atomic physics coexisting with a classical interpretation where a naked singularity remains.
 
Lastly, if the quantization we have described is picked by Nature for some $\theta$, one might speculate about the late stages of Hawking radiation on this quantum shell background. Since $m$ is a constant of the background, let us assume the black hole starts at a positive eigenvalue $E\lessapprox m$ for a given large $m$. As the black hole radiates, it would make transitions to successively lower levels, ending ultimately in the negative energy ground state. At this point Hawking radiation would stop since there is no lower state.  The closest physical picture corresponding to this is the negative mass Schwarzschild  metric. Therefore, if this scenario is realized, the endpoint of Hawking radiation would correspond to a naked singularity. 

However, if $m$ shrinks during Hawking radiation, the spectrum would be confined to the contracting band $(-m,m)$, and as the black hole reaches the Planck scale $m\approx 1$, the ground state would become positive, and ultimately go to zero as  $m\rightarrow 0$. This would support the common view that the endpoint of Hawking radiation is flat spacetime. Which of these scenarios occurs depends on how a scalar field couples to this quantum background. 

This is one of the topics of further study suggested by this work. Another is the comparison of this work with the black hole to white hole transitions speculated to arise in the loop quantum gravity quantization of the null shell \cite{Haggard:2014rza}. 

 \smallskip

\noindent{\it Acknowledgements:} This work was supported by NSERC of Canada. J.Z. and S.M.H were supported respectively by an AARMS Postdoctoral Fellowship and a Lewis Doctoral Fellowship. We thank Edward Wilson-Ewing,   Sanjeev Seahra, and  Jorma Louko for helpful discussions.

\bibliography{fermi-shell}

\end{document}